\documentstyle[aps,prl,epsfig,multicol]{revtex}

\tighten
\begin{document}
\draft
\title{Zeeman splitting of zero-bias anomaly in Luttinger liquids}
\author{A.V.~Shytov$^{1,2}$, L.I.~Glazman$^3$, and O.~A.~Starykh$^4$}
\address{$^1$~Institute for Theoretical Physics, University of California,
  Santa Barbara, CA 93106-4030 \\
  $^2$~L.D.~Landau Institute for Theoretical Physics, 
       2 Kosygina Str., Moscow, Russia 117940 \\
  $^3$~Theoretical Physics Institute, University of Minnesota,
  Minneapolis, MN 55455 \\
  $^4$~Department of Physics and Astronomy, Hofstra University, 
  Hempstead, NY 11549
}
\maketitle
\begin{abstract}
   Tunnelling density of states (DoS) in Luttinger liquid has a dip 
   at zero energy, commonly known as the zero-bias anomaly (ZBA). In
   the presence of a magnetic field, in addition to
   the zero-bias anomaly, the DoS develops two peaks
   separated from the origin by the Zeeman energy. We find the shape
   of these peaks at arbitrary strength of the electron-electron interaction.
   The developed theory is applicable to various kinds of
   quantum wires, including carbon nanotubes.
\end{abstract}

\begin{multicols}{2} 

Interaction between electrons in a conductor leads to the formation of
an anomaly in the tunnelling density of states (DoS) at the Fermi
level. The one-particle DoS is
directly related to the  differential conductance of a tunnel
junction, and the anomaly in DoS translates to the zero-bias anomaly (ZBA)
of the tunneling conductance.  
This anomaly gets stronger if the conductor is disordered, and
if the  dimensionality of the electron system is reduced. The
perturbative  treatment of the DoS anomaly in disordered conductors
is  well-developed~\cite{AAreview}. 
In a disordered wire or film,
the perturbation theory in the interaction strength is divergent at
the Fermi level, and therefore a non-perturbative treatment is needed 
to describe the DoS at low energies~\cite{Finkelstein,environment,KA}. 
In one-dimensional conductors  with one or a few propagating electron
modes the suppression of the DoS due  to
the electron-electron repulsion is strong even in the absence of
disorder.  The density of states in this case is adequately described
within the Luttinger liquid theory~\cite{KF}. 
The ZBA was observed in experiments with higher-dimensional disordered
systems~\cite{Imry,Dynes}. The recently 
measured~\cite{nanotubes-ll,intramolecular} strong
suppression of the tunnelling in a single-wall carbon
nanotube at low bias gave an evidence  that electrons in a nanotube
indeed form a Luttinger liquid.

Zero bias anomaly thus provides important information about
strongly correlated electron systems. However, ZBA is
sensitive mostly to the dynamics of electron charge but not to that of spin.
To probe the spin physics, one may study the effect of a magnetic
field on the properties of the electron system. The
perturbative calculation shows~\cite{AAreview} that the application of a
magnetic field  modifies the anomaly in the DoS. It acquires, in
addition to the zero-bias dip, two peaks at energies
$\varepsilon=\pm g\mu_BB$, where $g\mu_BB$ is the Zeeman energy.
The peaks heights are equal and proportional to the
electron-electron  interaction constant in the triplet
channel~\cite{AAreview}, which is not accessible in a measurement of
the conventional ZBA. The described Zeeman splitting of the ZBA in disordered
normal conductors was not observed yet, but a related phenomenon was
studied in thin superconducting films~\cite{Adams}. 

The perturbative theory of Zeeman splitting of ZBA in a clean 1D system was
developed in~\cite{YMG}. There, the anomalies in the 
conductance were ascribed to the physics of Bragg reflection of 
electrons off the Friedel oscillation. Magnetic field $B$ splits the
standard Friedel oscillation in two. The difference between the two
corresponding wave vectors is proportional to $B$. Electron
scattering off the two components of the Friedel oscillation results
in the conventional DoS anomaly at zero energy and two additional peaks in
the DoS at $\pm g\mu_B B$. This single-electron picture is
valid for a weak electron-electron interaction only,  and is not
applicable to the Luttinger liquid with a strong repulsion
between electrons. However, the strongest manifestation of the Luttinger
liquid behavior is found in carbon nanotubes, where the interaction is
not weak. Thus, one may question the existence of  peaks in the DoS at
Zeeman energy in such systems. 

In this paper, we demonstrate that the tunneling density of states 
in a Luttinger liquid is singular at energies
 $\varepsilon=\pm g^{\ast}\mu_B B$.
The effective Land\'e factor $g^{\ast}$ here is renormalized 
by the interaction.
The overall magnitude of  the singular correction  to DoS
is proportional to the  constant of electron-electron interaction 
in the triplet channel.
The energy dependence of the DoS around the singularities is given 
by a power law, 
$\delta\nu(\varepsilon) \sim |\varepsilon\pm g^{\ast}\mu_B B|^\gamma$.  
We calculate the exponent $\gamma$ in terms of the Luttinger liquid
parameters. 

Consider the one dimensional Luttinger liquid filling the 
half-line $x > 0$ and confined by a barrier at $x = 0$. 
We decompose the electron creation operator $\psi_s^{\dagger}(x)$
into the left- and right- moving parts: 
$\psi_s^{\dagger} = \psi_{+,s}^{\dagger} + \psi_{-,s}^{\dagger}$.  
Here $\pm$ denote left and right movers, and $s = \pm 1$ denote two
spin states.
Then, we bosonize electrons~\cite{Tsvelik}:
\begin{eqnarray}
\label{psi-operator}
\psi_{\pm,s}^{\dagger} (x)  &=&  \frac{1}{\sqrt{2\pi a}} 
    \exp\left\{
           \pm i p_F x \pm i \frac{\pi}{2} 
\right.
\\
&-&  \left. \frac{i}{\sqrt{2}} 
                \left[
                     \pm (\varphi_\rho (x) + s \varphi_\sigma (x)) 
                      +  \vartheta_\rho (x) + s \vartheta_\sigma (x)
                \right]  
        \right\} 
\ . \nonumber 
\end{eqnarray}
Bosonic variables $\varphi_i$ and $\vartheta_i$ describe 
charge ($i=\rho$) and spin ($i=\sigma$) fluctuations,
and $a$ is the short-distance cutoff. 
The fields $\varphi_i$ and $\vartheta_j$ are canonically conjugate: 
$[\varphi_i(x), \vartheta_j (x')] = i \delta_{ij} \theta(x - x')$.  
Since the electron wave function is  zero at the barrier ($x = 0$), the fields 
$\varphi_i$ satisfy boundary condition:
\begin{equation}
\label{boundary-condition}
\varphi_\rho (0) = \varphi_\sigma (0) = 0 \ . 
\end{equation}
The Hamiltonian can be divided into four parts: 
\begin{equation}
\label{hamiltonian}
H = H_\rho + H_\sigma + H' + H_B\  , \\
\end{equation}
where the first two terms include the density-density  
interactions:
\begin{equation}
\label{quadratic}
H_i =\ \frac{u_i}{2\pi} \int dx 
\left[ 
     K_i           (\partial_x \vartheta_i )^2 
   + \frac{1}{K_i} (\partial_x \varphi_i   )^2 
\right]  \ , 
\nonumber
\end{equation}
and $H'$ represents spin-flip backscattering:
$$
\label{backscattering}
H' = \frac{2 g_\perp}{(2\pi a)^2} \int dx \cos\sqrt{8}\varphi_\sigma 
\ . 
$$
The last term in Eq.~(\ref{hamiltonian}) describes Zeeman splitting:
$$
H_B = - \frac{g\mu_B B}{2} \int dx\, \rho_{\rm spin}(x) =  
\frac{g\mu_B B}{2\pi \sqrt{2}} \int dx \partial_x \varphi_\sigma 
\ , 
$$
where $\rho_{\rm spin}(x)$ is spin density.
The parameters $u_i$, $K_i$
and $g_\perp$ can be expressed in terms of interaction  
potential $V(x)$, but here we treat them as
phenomenological constants. For free electrons, 
$K_\rho = K_\sigma = 1$, while for repulsive interaction 
$K_\rho < 1$ and $K_\sigma > 1$. Also, the bare
parameters $K_\sigma$ and $g_\perp$ are not independent. 
For $g_\perp \ll 1$, they are related as 
$K_\sigma \approx 1 + g_\perp/2\pi u_\sigma$.

For convenience, we absorb the magnetic field term $H_B$ into
quadratic part by shift
$\varphi_\sigma \to \varphi_\sigma + g \mu_B  B x K_\sigma/\sqrt{2}u_\sigma $.
Then, the backscattering term transforms into 
\begin{equation}
\label{backscattering-periodic} 
H' = \frac{2 g_\perp}{(2\pi a)^2}\int dx \cos(\sqrt{8}\varphi_\sigma + b x)
\ ,
\end{equation}
with $b = 2 g\mu_B  B K_\sigma / u_\sigma$. 

The Hamiltonian (\ref{hamiltonian}) decouples into charge 
and spin sectors. While the charge excitations do not interact and
their Hamiltonian $H_\rho$ is quadratic, the spin sector 
is described by the sine-Gordon model $H_\sigma + H'$.
In zero magnetic field the constant $g_\perp$ is  
renormalized at low energies~\cite{Tsvelik},
\begin{equation}
\label{renormalization}
g_\perp(D) = 
\frac{g_\perp(W)}{1 +  \frac{g_\perp(W)}{\pi v_F}
  \log\frac{W}{D}}\ , 
\end{equation}   
where $W$ is the initial, and $D$ is the running bandwidths, 
and $g_\perp(W)$ is the  ``bare'' interaction constant. 
The renormalization group (RG) flow occurs
along the line $K_\sigma \approx 1 + g_\perp / 2 \pi v_F$
towards the fixed point $K_\sigma^\ast = 1$, $g_\perp^\ast = 0$. 
Finite magnetic field does not affect the RG flow for energies
larger than $ g\mu_B B$. For smaller energies, $K_\sigma$
becomes essentially independent of $\varepsilon$, while $g_\perp$
flows toward zero~\cite{RG-H}. 
In this way, the non-linear term $H'$ is not relevant at large times, 
and can be treated as a perturbation. 

We are interested in the DoS at $\varepsilon \to g\mu_B B$. This allows
us to consider the Hamiltonian $H_\sigma + H'$ 
acting on the states within the energy band of the width $D$ 
somewhat exceeding $2 g \mu_B B$.
The constants $K_\sigma(D)$ and $g_\perp(D)$ in the
Hamiltonian~(\ref{hamiltonian}) should be renormalized 
according to Eq.~(\ref{renormalization}).

Before developing a rigorous calculation, we provide a hint, where the
singularity in the DoS may come from. Tunneling of an electron may be
viewed as spreading of charge and spin densities, which initially
at $t=0$ were formed near the barrier ($x=0$). Because of the spin-charge
separation, the two density perturbations propagate independently. 
The charge propagates freely, since the corresponding Hamiltonian $H_\rho$ in
is quadratic. The propagation of the spin
density is affected by the backscattering term
Eq.~(\ref{backscattering-periodic}). To demonstrate qualitatively its
effect, we expand $H'$ in $\varphi_\sigma$ to the second order and then
derive the linear equation of motion for the field $\varphi_\sigma$. The
first-order expansion term only shifts  by a small
amount the solution of that equation. 
The second-order term generates a
contribution $\propto g_\perp\cos (bx)\varphi_\sigma$ in the equation of
motion, and leads to the phenomenon of Bragg reflection with wave
vector $b/2$. As the result, the backscattered component of
$\varphi_\sigma$ oscillates with frequency $\omega_z=u_\sigma
b/2=K_\sigma g\mu_B B$. These oscillations give rise
to features in the DoS at energies $\varepsilon=\pm\omega_z$.

We start with retarded Green's function
$$
G_s^R(x, x', \varepsilon) =  -i 
\int\limits_{0}^{\infty}  dt e^{i\varepsilon t}
    \left\langle \left\{
        \psi_{s}^{\dagger}(x, t)\, , \psi_{s} (x', 0)
     \right\}
     \right\rangle 
$$
(here $\{\dots\}$ is anticommutator) and compute the tunneling density
of states as~\cite{zero-wf}
\begin{equation}
\label{nu-definition}
\nu(\varepsilon)  = \frac{1}{4\pi (ik_F)^2} \sum_{s} 
\, {\mathop{\rm Im}\nolimits} 
\left.\frac{\partial^2}{\partial x \partial x'}\right|_{x=x'=0} 
\, G^{R}_s (x, x', \varepsilon)\ .
\end{equation}
For slowly varying $\varphi_i(x)$ and $\vartheta_i(x)$, 
one may neglect their derivatives, and differentiate
only the factors $e^{\pm ik_F x}$ in the electron 
operators~(\ref{psi-operator}). 
Equation~(\ref{nu-definition}) can be rewritten as
\begin{eqnarray}
\label{dos}
\nu(\varepsilon) =  \frac{1}{2\pi^2 a}\,{\mathop{\rm Re}\nolimits} 
\int\limits_{0}^{\infty} &dt&  \,
\left[
          {\cal G}_{\rho}(t)  e^{ i\varepsilon t} 
+         {\cal G}_{\rho}(-t) e^{-i\varepsilon t}  
\right] \nonumber \\ 
&\times& 
\left[
    {\cal G}_{\sigma}(t) + {\cal G}_{\sigma}(-t) 
\right]      
\end{eqnarray}
where 
\begin{equation}
{\cal G}_{i} (t) = 
\left\langle \mathop{\rm T}\nolimits
   \exp\frac{i\vartheta_i(x=0,t)}{\sqrt{2}} 
   \exp\frac{-i\vartheta_i(x=0,0)}{\sqrt{2}}
\right\rangle
\label{k-definition}
\end{equation}
are time-ordered Green's functions of charge and spin, 
and ${\rm T}$ denotes time-ordering.

To compute these correlation functions  at $g_\perp = 0$, 
we express the fields $\varphi_i$ and $\vartheta_i$  in terms of 
bosonic eigenmodes $a_q$, $a_q^{\dagger}$
of the Hamiltonian~(\ref{quadratic}).
Because of the  boundary condition~(\ref{boundary-condition}),  
only odd modes contribute to $\varphi_i$: 
\begin{eqnarray}
\label{phi-expression}
\varphi_{i} (x, t) &=& 
  \sum_{q} 
    c_{q,i}
    \sin qx \, 
    \left(
         a_{q,i} e^{i u_i q t} + a_{q, i}^{\dagger} e^{-iu_i q t}
    \right)\ ,   
\\
\label{theta-expression}
\vartheta_i(x, t) &=& 
  \sum_{q} 
    \frac{c_{q,i}}{K_i}
    \cos qx\, 
        \frac{a_{q,i} e^{i u_i q t} - a_{q, i}^{\dagger} e^{-iu_i q t}}{i}
    \ . 
\nonumber
\end{eqnarray}
Here $c_{q,i} = e^{-qa/2}\,\sqrt{\pi K_i/q}$, and
the short-distance cutoff $a = u_\sigma / D$ is related
to the reduced bandwidth $D$. 
The summation in Eq.~(\ref{phi-expression}) 
involves wave vectors $q > L^{-1}$, 
where $L$ is the length of the system. 
One can compute the average in Eq.~(\ref{k-definition})
using the relations~\cite{gaussian} 
\begin{equation}
\label{hausdorf}
e^{A}\,e^{B} = e^{A+B}\,e^{\frac{1}{2}[A,B]} 
\quad {\rm and } \quad 
\langle e^{A} \rangle = e^{\frac{1}{2}\langle A^2 \rangle} 
\ , 
\end{equation}
valid for any operators $A$ and $B$ linear in $a_q$ and
$a_q^{\dagger}$. At zero temperature, the only non-zero average
is $\langle a_{q} a_q^{\dagger} \rangle = 1$, and one finds
\begin{equation}
\label{k-nonperturbed}
{\cal G}_i^{(0)} ( t) 
= \left(\frac{i a}{u_i |t| + ia}\right)^\frac{1}{2K_i} 
\ .
\end{equation}
Substituting this expression into Eq.~(\ref{dos})
and evaluating the integral for $\epsilon \ll D$,
one arrives to the well-known formula~\cite{KF},
\begin{equation}
\nu^{(0)}(\varepsilon) = 
\frac{C_0}{\Gamma (1 + \alpha)} 
\, |\varepsilon|^{\alpha} \  
\ ,
\end{equation} 
with the anomalous exponent 
$\alpha = (K_\rho^{-1} + K_\sigma^{-1})/2 - 1$, 
and the prefactor
$$
C_0 = \frac{1}{\pi a} 
\left(\frac{a}{u_\rho}\right)^{\frac{1}{2K_\rho}} 
\left(\frac{a}{u_\sigma}\right)^{\frac{1}{2K_\sigma}} \ . 
$$

To develop perturbation theory in $g_\perp$, we 
use the standard expression~\cite{many-body}
\begin{equation}
\label{correction-casual}
{\cal G}_\sigma (t) = 
\frac{\left\langle 
    \mathop{\rm T}\nolimits 
    \exp\left[\frac{i\vartheta_\sigma(0,t)}{\sqrt{2}}\right]
    \exp\left[\frac{-i\vartheta_\sigma(0,0)}{\sqrt{2}}\right]
    S
\right\rangle_0}{\langle S \rangle_0} \ , 
\end{equation}
where
the averaging is performed 
over the non-perturbed Hamiltonian $H_\sigma$. 
To the first order in $g_\perp$, the  S-matrix is  
$$
S = 1 - i \int\limits_{-\infty}^{\infty} H'(t')\, dt'  \ . 
$$
Computing the correction 
\begin{eqnarray}
&&\delta{\cal G}_\sigma(t) = - \frac{2 ig_\perp}{(2\pi a)^2}
\int\limits_{-\infty}^{\infty}\! dt'\!\int\limits_{0}^{\infty}\! dx
\left\langle {\mathop{\rm T}\nolimits}
\exp\left[\frac{i(\vartheta_\sigma(t)
-\vartheta_\sigma(0))}{\sqrt{2}}
\right]
\right.
\nonumber
\\
&&\times
\left.
\Big[
\cos(\sqrt{8}\phi_\sigma(x, t') + bx)
- \left\langle \cos(\sqrt{8}\phi_\sigma(x, t') + bx)  \right\rangle_0
\Big] 
\right\rangle_0
\nonumber
\ , 
\end{eqnarray}
we use Eqs.~(\ref{phi-expression}) and~(\ref{hausdorf}), and arrive to 
\begin{eqnarray}
\delta {\cal G}_\sigma(t)
&=&   - \frac{2 i g_\perp}{(2\pi a)^2} \, 
\int\limits_{-\infty}^{\infty} dt' 
\int\limits_{0}^{\infty} dx \left(\frac{a^2}{a^2 + 4x^2}\right)^{K_\sigma}
\\
&\times& 
{\cal G}_\sigma^{(0)}(t)\,
\left[ \delta{\cal G}_{+}(x, t, t')\, e^{ibx} +  
       \delta{\cal G}_{-}(x, t, t')\, e^{-ibx} 
\right]
\ , \nonumber  
\end{eqnarray}
where 
\begin{eqnarray}
\label{correction-integrand}
\delta{\cal G}_{\pm} (x, t, t') &=& 
\pm \frac{2x u_\sigma t}{u_\sigma(t - t') \mp x 
                         + ia\, \mathop{\rm sgn}\nolimits (t-t')} 
\\
&\times& \frac{1}{u_\sigma t' \mp x 
                  + ia\, \mathop{\rm sgn}\nolimits (t')}
\ . \nonumber
\end{eqnarray}
Unlike Eq.~(\ref{k-nonperturbed}), the correction $\delta{\cal G}_\sigma(t)$
contains an oscillating part. We will retain only this part, 
since we are interested in singularities at non-zero
energies.
The oscillation originates from the point of enhanced 
singularity in Eq.~(\ref{correction-integrand}),
$x = u_\sigma t'  = u_\sigma t / 2$. 
The oscillating contribution to $\delta {\cal G}_\sigma(t)$ is
\begin{equation}
\frac{\delta {\cal G}_\sigma(t)}{{\cal G}^{(0)}_\sigma(t)}  = 
- \frac{i g_\perp}{2 u_\sigma} \,  
\left(
\frac{a}{u_\sigma |t|}
\right)^{2(K_\sigma - 1)}
\,  \theta (t) \,  e^{i \omega_z t}\,
\ ,   
\label{k-sigma}
\end{equation}
with $\omega_z = g^{\ast} \mu_B B$, and renormalized Land\'e factor
$g^{\ast} = K_\sigma g$. 
Using Eqs.~(\ref{dos}) and~(\ref{k-nonperturbed}) at $t > D^{-1}$, 
one finds the correction to the DoS
\begin{equation}
\label{int-correction}
\delta \nu(\varepsilon) = \pi C_0 
\mathop{\rm Re}\nolimits 
e^{i \frac{\pi}{2} (\alpha + 1)}
\int\limits_{D^{-1}}^{\infty} 
\frac{\cos \varepsilon t\,  dt}{t^{\alpha + 1}}\,  
\frac{\delta {\cal G}_\sigma (t)}{{\cal G}^{(0)}_\sigma(t)} \ ,
\end{equation} 
which is singular at $\epsilon = \pm\omega_z$.
Since Eq.~(\ref{int-correction}) is even in $\varepsilon$, we consider
further only $\varepsilon\approx\omega_z$. Integrating over the time domain
$t\sim |\varepsilon - \omega_z|^{-1}$, we find for the singular part:
\begin{eqnarray}
\label{result}
\frac{\delta \nu(\varepsilon)}{\nu^{(0)}(\omega_z)} &=& 
- \frac{g_\perp (\omega_z)}{4 u_\sigma} 
\frac{1}{\sin \pi \gamma}
\left|\frac{\varepsilon - \omega_z}{\omega_z}\right|^{\gamma}
\\
&\times&
\frac{\Gamma(1 + \alpha)}{\Gamma(1 + \gamma)}
\times 
\left\{
\begin{array}{lcl}
\cos\frac{\pi}{2}(\alpha - \gamma) & {\rm for }\  & \varepsilon > \omega_z \\
\cos\frac{\pi}{2}(\alpha + \gamma) & \            & \varepsilon < \omega_z
\end{array}
\right.
\ ,  
\nonumber
\end{eqnarray}
with the exponent $\gamma = \alpha + 2(K_\sigma - 1)$, {\it i.~e.}
\begin{equation}
\label{exponent}
\gamma = \frac{1}{2} \left(K_\rho^{-1} + K_\sigma^{-1} - 2\right)
+ 2(K_\sigma - 1) \ . 
\end{equation}
Equation~(\ref{result}) is valid for arbitrarily strong interaction
in the charge channel, and confirms the existence of singularity in DoS 
centered at energy
\begin{equation}
\omega_z= g^{\ast} \mu_B B \ ; \quad 
\frac{g^{\ast}}{g}=1+\frac{g_\perp(D\sim \mu_B  B)}{2\pi u_\sigma}
\ . 
\label{position}
\end{equation} 
At small $g_\perp$, which implies $K_\sigma \approx 1$, 
the main contribution to the exponent in 
Eq.~(\ref{result}) comes from the charge mode. Therefore, the exponents
$\alpha$ and $\gamma$ of the power-law singularities in the DoS 
at $\varepsilon=0$ 
and $\varepsilon=\omega_z$ respectively, are nearly identical, 
$\gamma \approx \alpha$. 
The contribution~(\ref{result}) 
was found for zero temperature, and 
its energy dependence is non-analytic at any $\gamma$. 
However, it may be easily distinguished from the regular part of
$\nu(\varepsilon)$ only  at $\gamma<1$.  Also, finite temperature $T$ smears
the singularity at $|\epsilon -\omega_z| \simeq T$. 

The DoS anomaly~(\ref{result}) is directly related to the bias dependence of
the tunneling conductance $G(V)$ between a conventional metal and a 
one-dimensional conductor. 
The corresponding singular contributions are related as
$\delta G(V)/G=\delta\nu(eV)/\nu$.
Tunneling between the ends of two identical one-dimensional 
conductors (such as
an intramolecular junction 
between carbon nanotubes~\cite{intramolecular}) also has a
singularity at Zeeman energy. In this case the peak 
at $eV = \omega_z$ has a different shape, because it is defined by the
singularities in the DoS both at $\varepsilon=\omega_z$ and $\varepsilon=0$. 
The conductance can be calculated as $G(V)=dI/dV$, where the tunneling
current $I(V)$ between the two conductors is proportional to the
convolution of the two corresponding DoS:
\begin{equation}
I(V) \propto \int\limits_{0}^{eV} d\varepsilon\,  
\nu(\varepsilon)\, \nu(\varepsilon - eV)\ . 
\end{equation}
Calculating this integral, one finds the singular contribution
to the differential conductance
\begin{eqnarray}
\label{doubled-conductance}
\frac{\delta G(V)}{G(V)} &=& 
- \frac{g_\perp}{ 4 u_\sigma}  
\frac{1}{\sin\frac{\pi}{2}(\alpha + \gamma)}
\,\left|\frac{eV - \omega_z}{\omega_z}\right|^{\alpha + \gamma}
\, 
\\
&\times&
\frac{\Gamma(1 + 2 \alpha)}{\Gamma(1 + \alpha + \gamma)} 
\nonumber 
\ . 
\end{eqnarray}
The exponent $\alpha + \gamma$ here coincides with $2\gamma$
up to a small  term of the order of $K_\sigma - 1$. 
This ``exponent doubling'' at $eV = \omega_z$ is similar to 
that occuring at zero bias~\cite{intramolecular}. 

It is interesting to analyse Eq.~(\ref{doubled-conductance}) 
in the limit of weak interactions, in which~\cite{Tsvelik}
$$
K_\sigma - 1\approx \frac{g_\perp}{2\pi v_F} \approx \frac{U(2k_F)}{2\pi
  v_F}
\ , \quad 
K_\rho \approx K_\sigma - \frac{U(0)}{\pi v_F}
\ .
$$
To the first order in the interaction potential $U(q)$, 
Eq.~(\ref{doubled-conductance})  yields 
$\delta
G/G=(U(2k_F)/4\pi u_\sigma)\ln(|eV-\omega_z|/\omega_z)$. 
This result is in agreement
with the first-order expansion of the tunneling conductance obtained
in~\cite{YMG}. However, beyond this order, there is a difference between
Eq.~(\ref{doubled-conductance}) and  Eq.~(47) 
in~\cite{YMG}. 
It stems apparently from the inapplicability of the RG approach developed
in~\cite{YMG} for the treatment of tunneling at energies close to $\omega_z$.

We derived Eq.~(\ref{result}) and~(\ref{exponent}) for the single-mode
Luttinger liquid. In the case of carbon nanotubes, one has to 
take into account the degeneracy between the two 
conic points in the Brillouin zone~\cite{bandstruct}. 
Treating the interaction in the charge channel non-perturbatively, 
as we did before, 
we find the exponent of the  singularity at $\epsilon = \omega_z$,  
\begin{equation}
\gamma_{\rm nt} \approx \alpha_{\rm nt} 
\approx \frac{1}{4}\left(K_\rho^{-1} - 1\right)\ . 
\end{equation}
Thus, the exponent $\gamma_{\rm nt}$ 
again nearly coincides with the ZBA exponent $\alpha_{\rm nt}$.
The reported values of $\alpha_{\rm nt}$ in the experiments 
with carbon nanotubes were  $\alpha_{\rm nt}=0.3 \div 0.6$, 
and therefore the peak at $\epsilon = \omega_z$
should be sharp and easy to observe. 

To conclude, the  application of a magnetic field to a
Luttinger liquid creates additional singularities (two peaks) in the tunneling 
density of states. 
We have found the power law characterizing these singularities  
and demonstrated that they are  robust, {\it i. e.}, they persist at any 
interaction strength in the charge channel. 
The magnitude of the peaks is
determined by the short-range interaction, which plays a minor role in
the charge physics of a Luttinger liquid and therefore is
hardly accessible in the measurements of the conventional zero-bias
anomaly of the tunneling conductance.

We are grateful to L.S.~Levitov, P.B.~Wiegmann, K.A.~Matveev,  
M.P.A.~Fisher, A.G.~Abanov and I.L.~Aleiner
for useful discussions.  This research was supported by NSF grants
PHY99-07949, DMR97-31756, DMR02-37296, EIA02-10736, KITP Scholarship and 
by an award from the Research Corporation.

\end{multicols}


\begin{references}
\bibitem{zero-wf} The electron wave function is zero at the barrier, and one
should therefore consider tunneling in a small vicinity of the
barrier. Thus, the tunneling operator contains derivatives of
the electron operator.
\end{references}
\end{document}